\documentstyle[11pt,aaspp4]{article}

\def\ltsima{$\;\buildrel < \over \sim \;$}
\def\simlt{\lower.5ex \hbox{\ltsima}}
\def\gtsima{$\;\buildrel > \over \sim \;$}
\def\simgt{\lower.5ex \hbox{\gtsima}}
\begin{document}
\title{ISO observations of
molecular hydrogen in HH 54$\bf ^*$:\\
measurement of a non-equilibrium ortho- to para-H$_2$
ratio}

\author{David A. Neufeld$^1$, Gary J. Melnick$^2$
and Martin Harwit$^3$}

\vskip 0.5 true in \parskip 6pt
  
\noindent{$^1$ Department of Physics \& Astronomy,  The Johns Hopkins University, 3400
North Charles Street,  Baltimore, MD 21218}
   
\noindent{$^2$ Harvard-Smithsonian Center for Astrophysics, 60 Garden Street, Cambridge, MA 02138}

\noindent{$^3$ 511 H Street S.W., Washington, DC 20024--2725; also Cornell University}

\vskip 0.1 true in
{\parskip 6pt

\vskip 0.2 true in
\noindent{$^*$ Based on observations with ISO, an ESA project with instruments
     funded by ESA Member States (especially the PI countries: France,
     Germany, the Netherlands and the United Kingdom) with the
     participation of ISAS and NASA.}
\vskip 0.5 true in}


\begin{abstract}

We have detected the S(1), S(2), S(3), S(4) and S(5) 
pure rotational
lines of molecular hydrogen toward the outflow source HH 54, using
the Short Wavelength Spectrometer (SWS) on board the Infrared
Space Observatory (ISO).  The observed H$_2$ line ratios indicate the
presence of warm molecular gas with an H$_2$ density of at least 
10$^5 \,\rm cm^{-3}$ and a temperature $\sim 650$~K in which the
ortho- to para-H$_2$ ratio is only $1.2 \pm 0.4$, significantly
smaller than the equilibrium ratio of 3 expected in gas at that 
temperature.  These observations imply that the measured
ortho- to para-H$_2$ ratio is the legacy of an earlier stage in 
the thermal history of the gas when the gas had reached equilibrium 
at a temperature $\simlt 90$~K.  Based upon the expected
timescale for equilibration, we argue that the non-equilibrium
ortho- to para-H$_2$ ratio observed in HH 54 serves as a
chronometer that places a conservative  upper limit of $\sim 5000$~yr 
on the period for which the emitting gas has been warm.  The 
S(2)/S(1) and S(3)/S(1) H$_2$ line ratios measured toward HH 54 are
consistent with recent theoretical models of Timmermann for the conversion
of para- to ortho-H$_2$ behind slow, `C'-type shocks, but only if
the preshock ortho- to para-H$_2$ ratio was $\simlt 0.2$.

\end{abstract}

\section{Introduction}

Molecular hydrogen is the dominant constituent of dense interstellar 
gas, and H$_2$ quadrupole transitions are a significant coolant
of such gas over a wide range of temperatures and densities
(Neufeld, Lepp \& Melnick 1995).
Although quadrupole vibrational transitions of H$_2$ near $2 \,\mu$m
have been widely observed 
from interstellar clouds that are strongly irradiated
by ultraviolet radiation or have been heated by shock waves to temperatures 
$T \sim 2000$~K (e.g. Shull \& Beckwith 1982;
Beckwith et al.\ 1978; Gatley et al.\ 1987; Chrysostoniou et al.\
1993; and many others),  molecular hydrogen within cooler 
clouds that are not UV-irradiated is only 
detectable by means of its pure rotational lines.   With a broad
spectral coverage that is uninterrupted by atmospheric absorption regions, 
the Short Wavelength Spectrometer (SWS; de Graauw et al.\ 1996) 
on board the Infrared Space 
Observatory (ISO; Kessler et al.\ 1996) 
has allowed the complete H$_2$ rotational spectrum
to be observed from the interstellar medium for the first time.

Pure rotational emissions from H$_2$ have been detected by ISO toward
several regions of star formation within the Galaxy -- including S140 
(Timmermann et al.\ 1996), Cepheus A West (Wright et al.\ 1996),
the BD+40$^\circ$4124 group (Wesselius et al.\ 1996), and Orion-IRc2
(van Dishoeck et al.\ 1998) -- as well as towards
several external galaxies (e.g. Arp 220; Sturm et al.\ 1996).   
The observed H$_2$ line ratios provide valuable constraints upon the physical
conditions in the emitting gas, implying typical temperatures of 
500 - 800 K for those Galactic sources that have been observed to date.
These temperatures are much greater than those typical of cold
quiescent clouds ($\sim 30$~K), and presumably reflect the effects of
shock heating or radiative heating by a nearby star or protostar.
As far as we are aware, all ISO observations of such sources 
(prior to those reported in this Letter) are consistent with an ortho- 
to para-H$_2$ ratio of three, the value expected at temperatures
above $\sim 200$~K
if the observed regions 
have been warm long enough to reach equilibrium between the ortho 
(odd $J$ states with total nuclear spin 1) and para 
states (even $J$ states with total nuclear spin zero).\footnote{Other 
earlier astronomical observations of H$_2$ also
indicated ortho- to para-H$_2$ ratios consistent
with the values expected in thermal equilibrium.  Thus infrared
absorption line observations of NGC 2024 IRS 2 (Lacy et al.\ 1994)
implied an ortho- to para-H$_2$ ratio $< 0.8$, consistent with
the value of 0.2 expected at the temperature ($\sim 45$~K) of the 
absorbing gas, while ultraviolet absorption line observations
toward hot stars (Savage et al.\ 1977) using the Copernicus satellite
have revealed typical
ortho- to para-H$_2$ ratios $\sim 1$ in diffuse clouds, 
again consistent with thermal equilibrium at the 
temperatures $\sim 80$~K that are typical of such clouds.
We note that {\it vibrational} 
emissions from warm photodissociation
regions  and planetary nebulae often exhibit ortho- to
para-H$_2$ ratios {\it in excited vibrational states}
that are smaller than the value of 3 expected in equilbrium
 at high temperature 
(e.g. Hasegawa et al.\ 1987,
Ramsay et al.\ 1993, Chrysostomou et al.\ 1993, Hora \& Latter 1996), 
with typical
values in the range 1.7 - 2.1.  However, Sternberg \& Neufeld (1998)
have argued that these ortho- to para-H$_2$ ratios measured for excited
vibrational states do not require {\it true} 
ortho- to para-H$_2$ ratios smaller than 3
but are simply a consequence of optical depth effects in the fluorescent
pumping of the observed vibrational emissions.}

In this Letter we report the results of ISO observations of molecular 
hydrogen in the source HH 54, a Herbig-Haro object located
at the edge of the Chamaleon II dark cloud at
an estimated distance $\sim 200$~pc from the Sun
(Hughes \& Hartigan 1992).  The source of the outflow responsible
for HH 54 is believed to be IRAS 12496-7650 (Hughes et al.\ 1989),
a deeply embedded young stellar object lying approximately 
$4^\prime$ to the southwest. 
Previous observations
of infrared H$_2$ vibrational lines (Gredel 1994) and optical line emissions
(Schwarz \& Dopita 1980) have suggested the presence
of hot, shock-excited gas in HH 54. 

Our observations are described in 
\S 2 below, and our results reported in \S 3 below.   A discussion
follows in \S 4, in which particular emphasis is placed upon
the non-equilibrium
ortho- to para-H$_2$ ratio $\sim 1.2$ measured in HH 54.

\section{Observations and data reduction}

Using the SWS of ISO in its grating mode (SWS02), 
we observed the
S(1), S(2), S(3), S(4) and S(5) lines of H$_2$ toward the source 
HH 54 on 1997 November 4th.   The ISO beam was centered midway between
two strong sources of H$_2$ vibrational emission,
positions E and K (Sandell et al.\ 1987), at coordinates 
$\rm \alpha= 12h\,55m\,53.4s$, 
$\delta=-76^\circ \,56^\prime\,20.5^{\prime\prime}$ (J2000),
with the long axis of the beam oriented at position angle 118$^\circ$.
The total observing time on target was 6400 s, including overheads
for dark current measurements and calibration.
The rest wavelength, spectral resolution, and beam size are given 
in Table 1 for each of the five observed H$_2$ lines.

{\vskip 0.4 true in
\centerline{TABLE 1}
\centerline{H$_2$ rotational lines observed toward HH54 E+K}
\vskip 0.2 true in
{\doublespace
\scriptsize \begin{tabular}{c c c c c c c}  \hline 
Line & Wavelength & Velocity       & Beam size  & Line Flux$^{b}$ & Line Intensity$^{bc}$ & Upper state \\ 
     &            & Resolution$^a$ &          &                 &                       & column density$^{cd}$ \\
     & ($\mu$m)   & (km s$^{-1}$)  & (arcsec) & (W cm$^{-2}$)   &
($\rm erg \, cm^{-2} \, s^{-1} \, sr^{-1}$) 
& (cm$^{-2}$) \\ \hline 

S(1) & 17.0348 & 160  &  $14 \times 27$ &   $4.8 \times 10^{-20}$ & $5.4 \times 10^{-5}$ & $1.3 \times 10^{19}$ \\     
S(2) & 12.2786 & 240  &  $14 \times 27$ &   $2.1 \times 10^{-19}$ & $2.4 \times 10^{-4}$ & $7.5 \times 10^{18}$ \\
S(3) &  9.6649 & 160  &  $14 \times 20$ &   $1.7 \times 10^{-19}$ & $2.6 \times 10^{-4}$ & $2.2 \times 10^{18}$ \\ 
S(4) &  8.0251 & 200  &  $14 \times 20$ &   $1.6 \times 10^{-19}$ & $2.4 \times 10^{-4}$ & $5.0 \times 10^{17}$ \\  
S(5) &  6.9095 & 240  &  $14 \times 20$ &   $1.9 \times 10^{-19}$ & $2.9 \times 10^{-4}$ & $2.2 \times 10^{17}$ \\ 

\hline 
\\
\end{tabular}
\singlespace
\parskip 0pt

\noindent $^a$ FWHM of the instrumental profile for an extended source, from
the ISO SWS Observer's manual, Issue 3.0.  For a point source, 
the spectral resolution is roughly a factor of two better.

\noindent $^b$ uncorrected for extinction

\noindent $^c$ beam averaged

\noindent $^d$ computed with a correction for extinction (see text)

}}
\vfill\eject

The initial data reduction was carried out with version 6.22 of
the ISO pipeline software, and the ISAP 
software package\footnote{The ISO Spectral Analysis Package (ISAP)
is a joint development by the LWS and SWS Instrument Teams and
Data Centers.  Contributing institutes are CESR, IAS, IPAC, MPE,
RAL and SRON.} was then used to
remove bad data points and to co-add the individual spectral scans.

\section{Results}

Figure 1 shows the SWS H$_2$ spectra observed toward HH 54 E+K.  In every
case, the observed line width is no broader than the instrumental
response function, i.e.\ the lines are unresolved at the
resolution of the SWS grating mode.  

The measured line fluxes, beam-averaged line intensities and
beam-averaged H$_2$ column densities in rotational states $J=3$ through 7
are given in Table 1.  The current SWS flux calibration is believed accurate
to $\sim 30 \%$.
In computing the H$_2$ column densities, we used the spontaneous 
radiative rates of Turner, Kirby-Docken, \& Dalgarno
(1977; confirmed recently by the more precise
calculations of Wolniewicz, Simbotin \& Dalgarno 1998) 
for quadrupole transitions of H$_2$, which imply 
that the lines that we observed are all optically thin.
We also applied dust extinction corrections based upon the interstellar
extinction curves of Draine (1989), adopting Gredel's (1994)
estimate of 0.3 mag for $E(J-H)$ that was 
obtained from observations of the [FeII] lines
at 1.257 and 1.644~$\mu$m toward HH 54 E+K.  The
assumed extinctions for the S(1), S(2), S(3), S(4) and S(5)
lines were accordingly 0.10, 0.12, 0.30, 0.09 and 0.04 mag 
respectively\footnote{The assumed extinction for the S(3) line
--  and to a lesser extent for the S(1), S(2) and S(4) lines --
is enhanced by silicate features in the assumed extinction curve
(Draine 1989),  the strength of which may vary from one source to
another.  However, we found that the gas temperature and 
ortho- to para-H$_2$ ratio derived below are only very
weakly dependent upon the exact shape of the assumed extinction
curve and are negligibly altered
even if the silicate features are assumed to be entirely absent.}.

\begin{figure}
\plotone{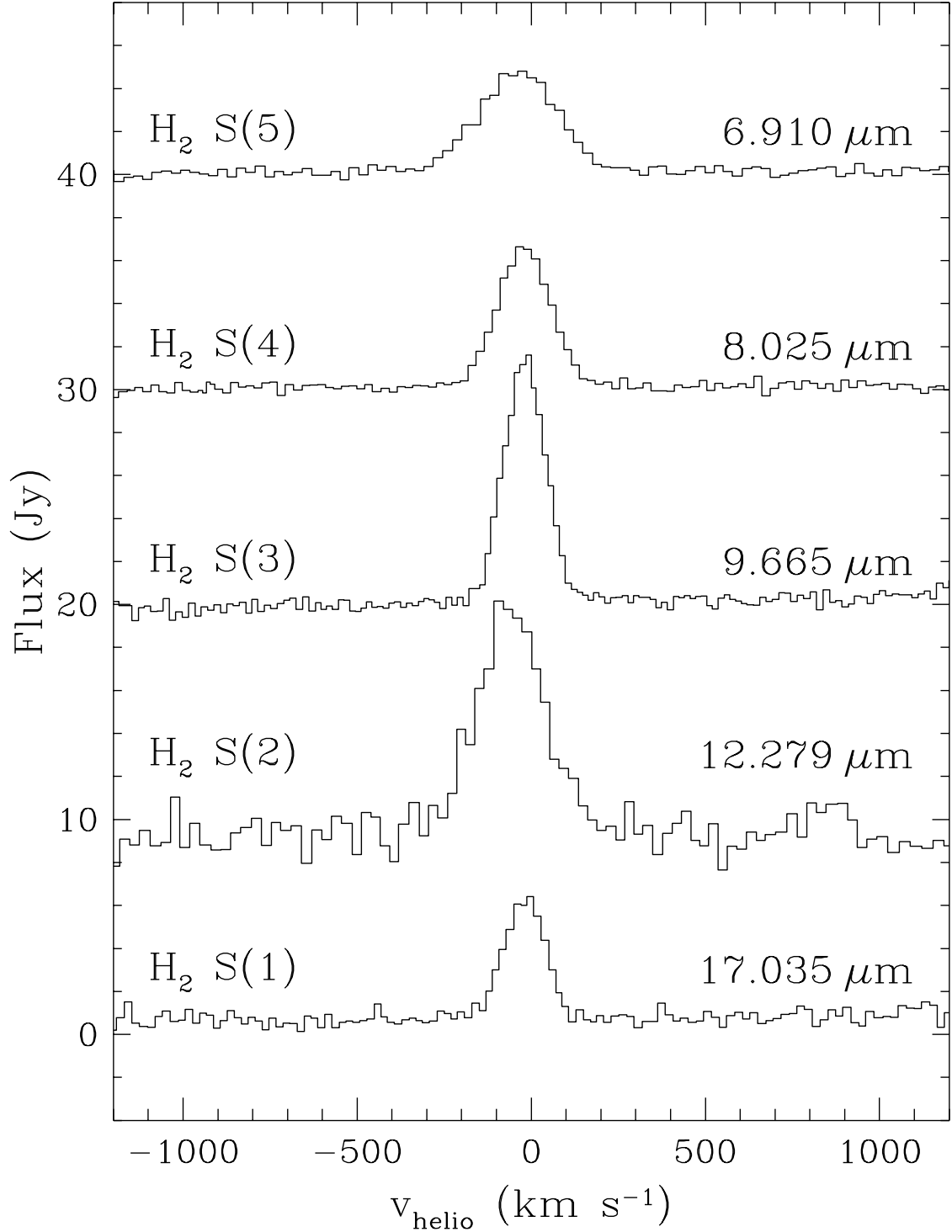}
\caption{ISO Short Wavelength Spectrometer (SWS) spectra of pure 
rotational lines of H$_2$ observed toward HH 54 E+K, for a beam of size
$14^{\prime \prime} \times 27^{\prime \prime}$ [S(1) and S(2) lines]
or $14^{\prime \prime} \times 20^{\prime \prime}$ [S(3), S(4)
and S(5)] centered at $\rm \alpha= 12h\,55m\,53.4s$, 
$\delta=-76^\circ \,56^\prime\,20.5^{\prime\prime}$ (J2000).
The S(2), S(3), S(4) and S(5) lines are respectively
offset by 10, 20, 30, and 40 Jy.}
\end{figure}

\begin{figure}
\plotone{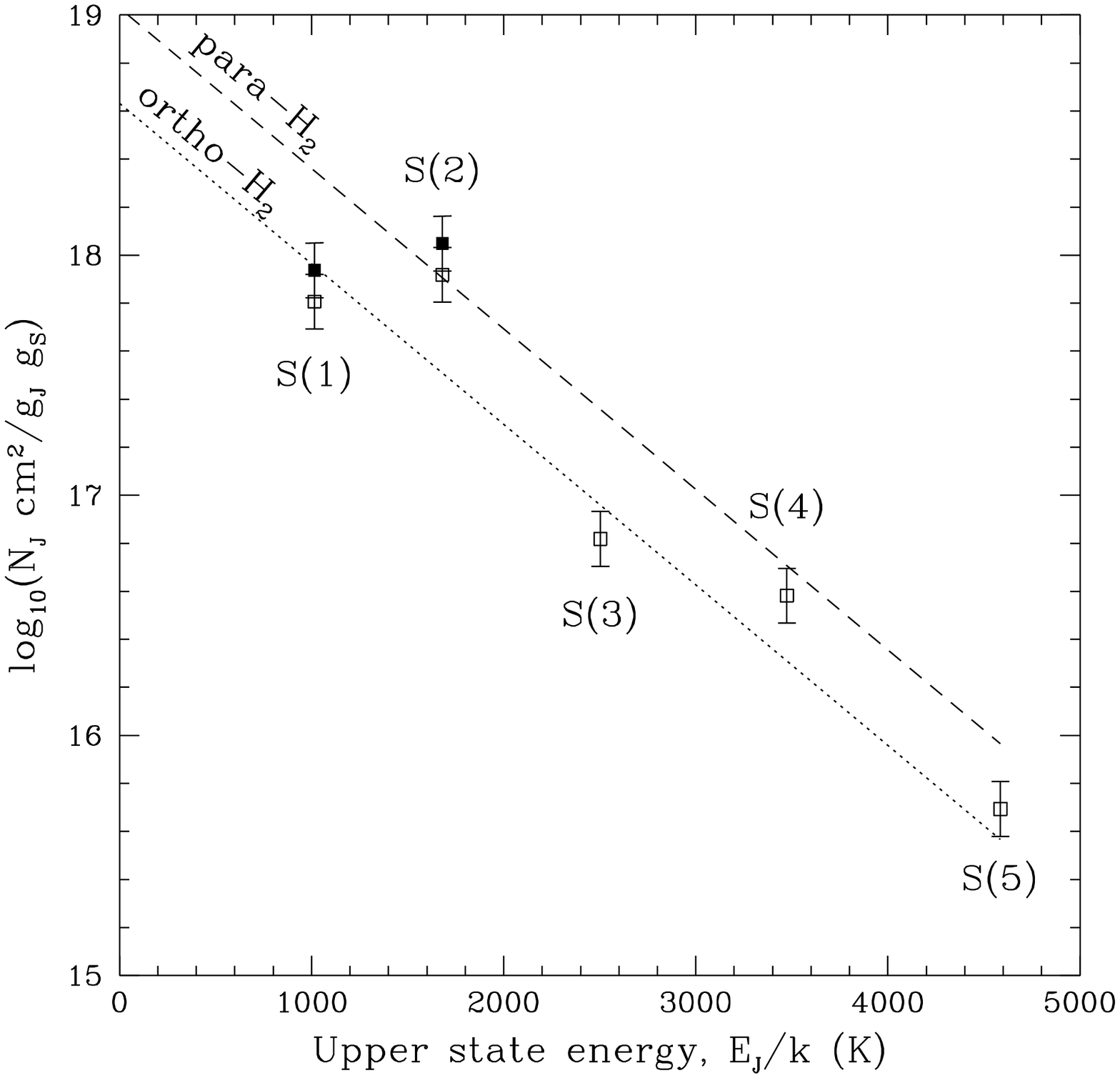}
\caption{Rotational plot for H$_2$ rotational states $J=3$ 
through 7.  The logarithm of $N_J/(g_J g_S)$ is plotted 
against $E_J/k$, 
where $N_J$ is the beam-averaged column density, $g_J= 2J+1$ is the 
rotational degeneracy, $g_S$ is the spin
degeneracy (1 for even $J$ and
3 for odd $J$)
and $E_J$ is the energy of the state of
rotational quantum number $J$.  
Open squares correspond to the values tabulated in Table 1,
with column densities averaged over regions of size 
$14^{\prime \prime} \times 27^{\prime \prime}$ for $J=3$ and 4 
and size $14^{\prime \prime} \times 20^{\prime \prime}$ for $J=5$, 6, and 7.  
Filled squares for $J=3$ and 4 show values that have
been corrected to the smaller beam size of 
$14^{\prime \prime} \times 20^{\prime \prime}$, 
on the assumption that the emission region is small compared to either 
beam.  Straight lines show the fit for a gas temperature 
of 650~K and an ortho- to para-H$_2$ ratio of 1.2 (see text).}
\end{figure}

Figure 2 shows a ``rotational plot'' for H$_2$ rotational 
states $J=3$ through 7.  Here the logarithm of $N_J/(g_J g_S)$ is plotted 
against $E_J/k$,
where $N_J$ is the beam-averaged column density, $g_J= 2J+1$ is the 
rotational degeneracy, $g_S$ is the spin
degeneracy (1 for even $J$ and 
3 for odd $J$),
and $E_J$ is the energy of the state of
rotational quantum number $J$.  
Open squares correspond to the values tabulated in Table 1,
with column densities averaged over regions of size 
$14^{\prime \prime} \times 27^{\prime \prime}$ for $J=3$ and 4 
and size $14^{\prime \prime} \times 20^{\prime \prime}$ for $J=5$, 6, and 7.  
The filled squares for $J=3$ and 4 show values that have
been corrected to the smaller beam size of 
$14^{\prime \prime} \times 20^{\prime \prime}$, 
on the assumption that the emission region is small compared to either 
beam.  That assumption, adopted hereafter in the  
discussion, is appropriate if the 
H$_2$ S(1) and S(2) emission follows the compact distribution
of the higher-excitation
vibrational emission mapped by Gredel (1994).   

Figure 2 shows that the points corresponding $J=3$, 5, and 7 are
colinear to within the observational uncertainties, consistent with
the emission expected from gas in thermal equilibrium at a temperature
$T \sim 700$~K.  However, the column densities in $J=2$ and 4
lie substantially above the best-fit line for the ortho (odd-$J$ states),
implying that the ortho- to para-H$_2$ ratio (OPR) is smaller than the high
temperature equilibrium value of 3 expected at 700~K.  Our best fit to
all the data points implies an OPR of $1.2 \pm 0.4$,
a gas temperature of $650$~K, and a total H$_2$ column density
of $9 \times 10^{19} \rm cm^{-2}$ (averaged over a 
$14^{\prime \prime} \times 20^{\prime \prime}$ beam). 

The rotational temperature derived for HH 54 E+K is larger than
the temperature $\sim 330$~K derived by Liseau et al.\ (1996)
from ISO Long Wavelength Spectrometer (LWS)
observations of CO transitions towards HH 54 B, a difference
that presumably results from the different pointing and
larger LWS beam size used by Liseau et al.  On the other
hand, the H$_2$ rotational temperature we derived towards HH 54 E+K
is {\it smaller} than the temperature of 2100 K derived by Gredel (1994)
from the observed ratio of the H$_2$ $v=1-0$ and $v=2-1$ vibrational bands.
Again, such a discrepancy is unsurprising and probably indicates that
a mixture of shock velocities and gas temperatures is present 
within the beam, with vibrational emissions selectively probing 
hotter gas associated with faster shocks.

\section{Discussion}

Our measurement of an OPR smaller than 3
in gas of temperature $\sim 650$~K is {\it prima facie} evidence
that the gas has
not been warm long enough to reach equilibrium between the
ortho and para states; thus the measured OPR $\sim 1.2$
is the legacy of an earlier stage in the thermal history of the gas 
when the OPR had reached equilibrium at a value $\simlt 1.2$ (corresponding
to a gas temperature $\simlt 90$~K).  

We can place an upper limit on the time period, $\tau$, for which the
emitting gas has been warm by estimating the timescale for
conversion from para- to ortho-H$_2$, $\tau_{conv}$.  
In media of low fractional
ionization, the 
conversion of para- to ortho-H$_2$ is dominated by reactive 
collisions with atomic hydrogen, for which the rate coefficient at
650~K is $1.0 \times 10^{-13}\,\rm cm^3\,s^{-1}$ (Tin\'e et al.\ 1997): 
thus $\tau_{conv}$ is given by $3000\,[{\rm 100\,cm}^{-3} / n({\rm H})]
\,\rm yr$, where $n({\rm H})$ is the density of hydrogen atoms.
(Even if other interconversion mechanisms are significant, this
is in any case an upper limit on $\tau_{conv}$.)   

Models for the chemistry of dense molecular gas (e.g.
Neufeld, Lepp \& Melnick 1995) predict that
once the temperature of the gas exceeds $\sim 300$~K, atomic 
oxygen will be rapidly converted to water, a prediction that has been
supported by recent observations of water toward warm shocked regions in
the Orion Molecular Cloud (Harwit et al.\ 1998).   The production of
water takes place by means of a series of two hydrogen atom 
abstraction reactions with H$_2$ and therefore leads to the 
production of two hydrogen atoms for each water molecule produced;
thus the atomic hydrogen density in recently-heated gas at temperature
650~K is at least twice the atomic oxygen density that was present in the
gas prior to its being heated.  Assuming an initial O abundance, 
$n({\rm O})/n({\rm H}_2)$ of $3.5 \times 10^{-4}$ (consistent with the
gas-phase oxygen and carbon abundances of Cardelli et al.\ 1995 and the
assumption that O accounts for most of the gas-phase oxygen not
bound as CO), this places an upper limit of
$5000 \,[{\rm 10^5\,cm}^{-3} / n({\rm H_2})]\,\rm yr$ on $\tau_{conv}$.
(This conservative upper limit still applies, of course, even if
atomic hydrogen is produced by other mechanisms in addition to the 
reaction of O with H$_2$ to form water).
Finally, the fact that the $J=7$ state is apparently thermalized (c.f.
Figure 2) implies a lower limit on the H$_2$ density of $\sim 
{\rm 10^5\,cm}^{-3}$,
the critical density at which the collisional de-excitation rate
for $J=7$ (Tin\'e et al.\ 1997)
is equal to the spontaneous radiative decay rate (Turner et al.\ 1977).
Thus we conclude that the emitting gas that we observed
in HH 54 has been warm for a period 
$\tau \simlt \tau_{conv} \simlt 3000\,[{\rm 100\,cm}^{-3} / n({\rm H})]
\,{\rm yr} \simlt 5000\, [{\rm 10^5\,cm}^{-3} / n({\rm H_2})]\,{\rm yr} 
\simlt 5000{\,\rm yr}$.

Transient heating by shock waves -- observed widely in Herbig-Haro objects --
provides a natural explanation of the non-equilibrium OPR that
we observed.  Indeed, a non-equilibrium OPR within molecular gas heated by
slow shocks has been predicted recently by Timmermann (1998), who 
presented a detailed theoretical treatment of ortho-para interconversion
within slow `C'-type shocks.  Timmermann obtained predictions for the 
H$_2$ rotational line strengths 
for shocks of velocity 10 to $30 \, \rm km \, s^{-1}$ that propagate in
gas of preshock H$_2$ density $5 \times 10^3$, $5 \times 10^4$ and 
$5 \times 10^5\,\rm cm^{-3}$
and preshock ortho to para-H$_2$ ratio OPR$_i = 1$ or 3.  The results 
show that para-to-ortho conversion is incomplete for shock
velocities smaller than 20 to 25 $\rm km\,
s^{-1}$ -- the timescale
for equilibration being longer than the shock timescale -- so that slow
shocks with OPR$_i = 1$ show a non-equilibrium OPR within the
postshock region.

\begin{figure}
\plotone{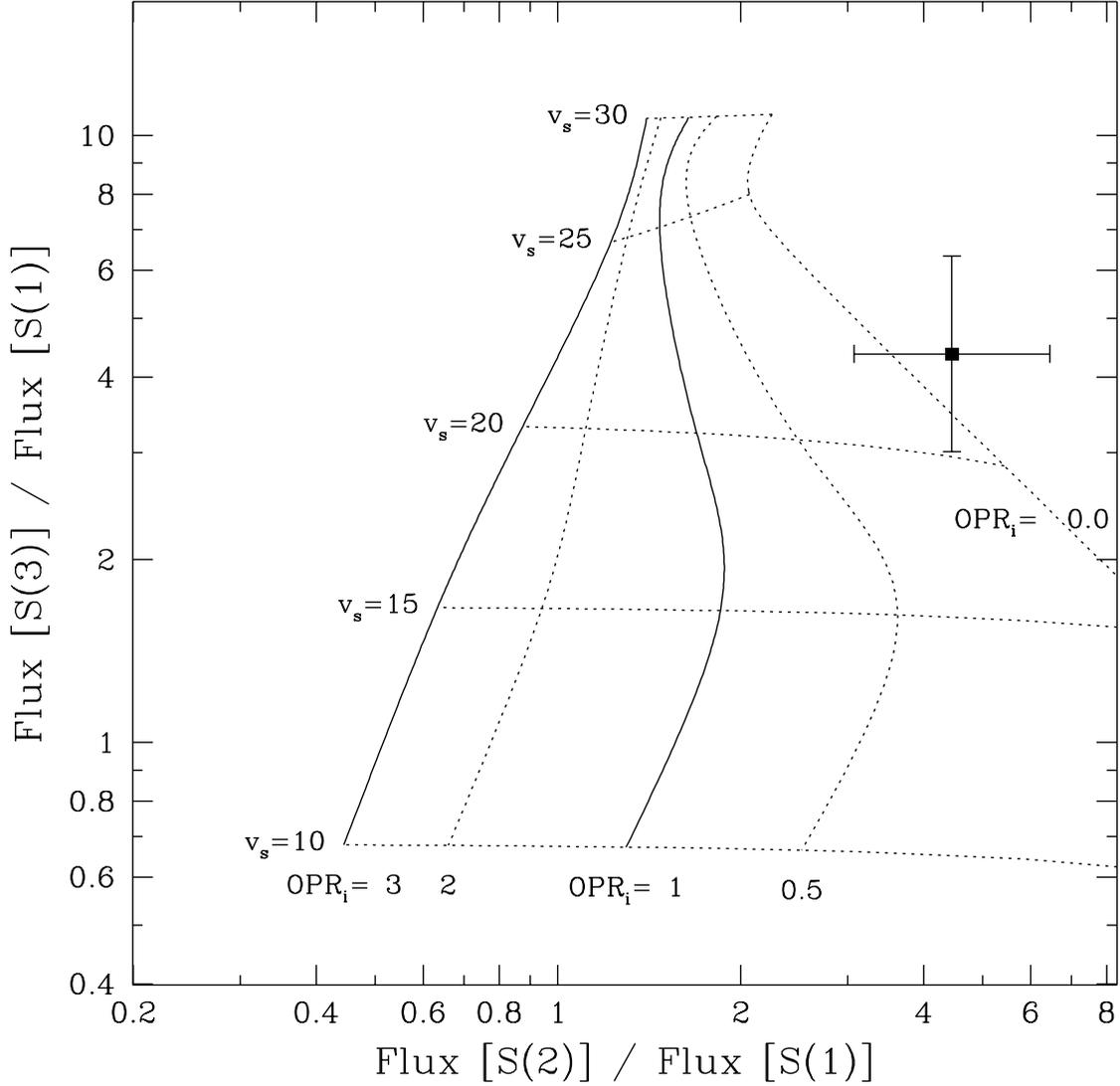}
\caption{Comparison between the observed
S(2)/S(1) and S(3)/S(1) H$_2$ line flux ratios (filled square)
and the predictions of Timmermann (1998)
for a `C'-type shock propagating in gas of H$_2$ density 
$5 \times 10^5 \,\rm cm^{-3}$,
shown as a function of the shock velocity, $v_s$, and
the initial (i.e.\ preshock) ortho- to para-H$_2$ ratio,
OPR$_i$.
The observed fluxes have been corrected for
extinction (see text).   
We have extrapolated 
or interpolated Timmermann's results
to cases (dotted lines) where the initial OPR is other
than 1 or 3, assuming
that the line fluxes depend linearly upon
the initial para-H$_2$ fraction, $({\rm OPR}_i+1)^{-1}$. }
\end{figure}

Figure 3 shows a quantitative comparison between the observed
S(2)/S(1) and S(3)/S(1) line flux ratios and Timmermann's predictions for
a preshock H$_2$ density of 
$5 \times 10^5 \,\rm cm^{-3}$.  In fact, none of the
shock parameters actually considered by Timmermann yields a S(2)/S(1) ratio
large enough (i.e.\ a postshock OPR small enough) to match the
observations.  However, we have interpolated and extrapolated 
Timmermann's results to cases (dotted lines) where the initial OPR is other
than 1 or 3, assuming
that the line fluxes depend linearly\footnote{A linear dependence
would obtain exactly provided that the temperature structure of
the shock is independent of the initial OPR.  Our
extrapolation method yields the expected result that the 
fluxes in ortho-H$_2$ transitions tend to zero in the limit of
small initial OPR and small shock velocity.}  upon
the initial para-H$_2$ fraction, $({\rm OPR}_i+1)^{-1}$.
Our extrapolation shows that 
a shock of velocity $\sim 22 \, \rm km \,s^{-1}$ can provide a satisfactory
fit to the S(2)/S(1) and S(3)/S(1) line ratios,
given a sufficiently small initial 
ortho- to para-H$_2$ ratio,
${\rm OPR}_i \simlt 0.2$, 

The results of Timmermann et al.\ (1998) also provide an explanation
for why previous studies (e.g. Smith, Davis \& Lioure 1997) of 
vibrational emissions from protostellar outflows and Herbig-Haro 
objects have always revealed an OPR close to 3: vibrational 
emissions inevitably trace faster, hotter shocks in which the
atomic hydrogen fraction is enhanced by streaming ion-neutral 
collisions and the rate coefficients for reactive collisions
between H and H$_2$ are larger.  Thus Timmermann's models
predict that  para-to-ortho conversion
is rapid enough to yield an equilibrium OPR of 3 behind any shock that
is fast enough to excite significant vibrational emission.

It is pleasure to acknowledge the support provided by
the Infrared Processing and Analysis Center (IPAC).
We also thank members of the help desk at Vilspa for their
assistance.  We gratefully acknowledge the support
of NASA grants NAG5-3316 to D.A.N.;
NAG5-3347 to M.H.; and NAG5-3542 and NASA contract NAS5-30702
to G.J.M.

\end{document}